\input harvmac
\settabs 3\columns \+&& SCIPP-03/05, RUNHETC-2003-15 \cr

 \vskip.4in \centerline{\bf  SUSY AND THE HOLOGRAPHIC SCREENS} \vskip.3in

\centerline{\it T. Banks}\vskip.1in \centerline{Department of
Physics and Institute for Particle Physics} \centerline{University
of California, Santa Cruz, CA 95064} \centerline{and}
\centerline{Department of Physics and Astronomy, NHETC}
\centerline{Rutgers University, Piscataway, NJ 08540} \vskip.3in
\centerline{E-mail: \ banks@scipp.ucsc.edu} \vskip.5in

\centerline{\bf ABSTRACT}

\noindent The Cartan-Penrose (CP) equation is interpreted as a
connection between a spinor at a point in spacetime, and a pair of
holographic screens on which the information at that point may be
projected.  Local SUSY is thus given a physical interpretation in
terms of the ambiguity of the choice of holographic screen
implicit in the work of Bousso.  The classical CP equation is
conformally invariant, but quantization introduces metrical
information via the
B(ekenstein)-H(awking)-F(ischler)-S(usskind)-B(ousso) connection
between area and entropy. A piece of the classical projective
invariance survives as the $(-1)^F$ operation of Fermi statistics.
I expand on a previously discussed formulation of quantum
cosmology, using the connection between SUSY and screens.
\vfill\eject

\def\hn{{\cal H}_n}

\newsec{\bf Introduction: The Importance of Being SUSic}

String/M-theory is our most ambitious attempt to create a
complete quantum theory of the world.  Its success thus far has
been a quite spectacular demonstration of the existence of
mathematically well defined quantum theories which include
Einstein's theory of gravity as a low energy limit. In fact, all
extant examples are based on supergravity.  Indeed, with the
advent of string dualities, it is reasonable to say that
String/M-theory is distinguishable from other hypothetical
theories of gravity merely by its insistence on incorporating
local SUSY into the formalism.  The assumption of quantum
supersymmetry in various compactifications of 11 dimensional
supergravity, predicts the existence moduli spaces of quantum
vacuum states, as well as BPS string states, whose tension scales
to zero in certain extreme limits of moduli space.  This suggests
a weakly coupled string description of those limiting regimes,
with the string coupling identified as a power of the ratio
between the string tension and the 11 dimensional Planck scale.
The predictions give us precisely the known moduli spaces of
perturbative string theories, with the exception of certain
asymmetric orbifold constructions.  This is surely the most
elegant way to understand the heterotic string.

One is led to ask whether there is a more geometrical way of
understanding the necessity for incorporating supersymmetry into
a quantum theory of gravity.  In my lecture at the Strings at the
Millenium Conference in January of 2000\ref\tbmill{T.~Banks, {\it
Supersymmetry and Space Time} Talk given at the Strings at the
Millenium Conference, CalTech-USC, January 5, 2000.} , I suggested
that the geometrical origin of local SUSY was the holographic
principle.  The formulation of the holographic principle for
asymptotically flat spacetimes\ref\thornsuss{C.~Thorn, {\it
Reformulating String Theory With The 1/N Expansion},
hep-th/9405069; L.~Susskind, {\it The World as a Hologram},
J.Math.Phys. 36 (1995) 6377, hep-th/9409089.} makes it apparent
that the choice of a holographic screen (holoscreen) is a gauge
choice. In extant formulations in asymptotically flat space-times,
the holoscreen is chosen to be a light plane, corresponding to a
particular choice of light front gauge.

Bousso's\ref\raph{R.~Bousso, {\it A Covariant Entropy Conjecture
}, JHEP 9907 (1999) 004, hep-th/9905177 ;{\it Holography in
General Space Times}, JHEP 9906 (1999) 028, hep-th/9906022; {\it
The Holographic Principle for General
Backgrounds},Class.Quant.Grav. 17 (2000) 997, hep-th/9911002.}
general formulation of the holographic principle makes it
abundantly clear that there are many ways to project the data in a
given spacetime into collections of holoscreens.  Anti-de Sitter
spacetimes are a very special case in which there is a natural
screen on the boundary of spacetime. Even here, one might want to
introduce alternative screens, in order to understand approximate
locality in the bulk.  Thus, Bousso's work suggests that any
quantum theory of gravity based on the holographic principle must
have a new gauge invariance, going beyond (but intertwined with)
general coordinate invariance.  The classical description of
string/M-theory does have such a new local invariance, namely
local SUSY.  It was natural to consider whether this is in fact
the holoscreen invariance suggested by most discussions of the
holographic principle.

The mathematical connection comes from the Cartan-Penrose (CP)
equation, relating a null direction to a pure spinor.  In fact, a
pure spinor not only defines a null direction, but also a
holoscreen transverse to that direction.  In the next section, we
will recall the geometrical interpretation of so-called
pure\foot{Cartan's original notion of purity is somewhat different
than that recently introduced by Berkowitz to quantize
supersymmetric branes.} spinors by Cartan and Penrose, and argue
that one can think of the choice of a pure spinor at each point
in space-time as a choice of a holographic screen on which the
data at that point is projected. A local change of the choice of
spinor corresponds to the holoscreen gauge invariance noted
above, and smells like it has something to do with local SUSY.

The spinors in the classical CP equation are bosonic, but
projective.  Nothing in the correspondence depends on the overall
complex scale of the spinor.   This should be viewed as a
classical gauge invariance.  The current,
$\bar{\psi}\gamma^{\mu}\psi$ constructed from the pure spinor, is
a null direction, again a projective object. The equation knows
only about the conformal, and not the metrical structure of
space-time.  In particular, although the classical CP equation
makes a natural connection between the choice of a pure spinor at
a point in spacetime, a null direction, and the {\it orientation}
of a holographic screen, it says nothing about how far the screen
is from the point, nor the geometrical extent of the screen.

The geometrical interpretation of pure spinors originated in the
work of Cartan\ref\cart{E.~Cartan, {\it The Theory of Spinors},
MIT Press, Cambridge MA, reprinted from the original French
edition, Hermann, Paris, 1966.}.  In four dimensions, Penrose
described the geometry with the picturesque phrase ``A spinor is a
flagpole (null direction) plus a flag
(holoscreen)."\ref\mtw{C.W.~Misner, K.S.~Thorne, J.A.~Wheeler,
{\it Gravitation}, 41.9, p. 1157, W.H.Freeman, San Francisco,
1973, and references therein.}.

Below we will turn the CP spinors into quantum operators.  The
quantization procedure breaks the classical projective invariance
of the CP equation, leaving over a discrete phase invariance. I
view this as a statement that the metrical structure of space-time
has its origin in quantum mechanics.  This connection is suggested
by the BHFSB relation between entropy and the area of holoscreens.
The residual discrete phase invariance left over by the
quantization of CP spinors, should be viewed as a gauge symmetry
of the quantum theory. We will show that, after performing a Klein
transformation using a $Z_2$ subgroup of this gauge symmetry, the
CP spinors become Fermions and $Z_2$ is just $(-1)^F$.

The motivation for this particular method of quantizing the
classical variables, combines the classical relation between
spinors and holoscreens with the BHFSB relation between area and
entropy.   In this way the breaking of projective invariance is
precisely equivalent to the introduction of metrical structure on
space-time.  The detailed answer to the question of the size and
location of the holoscreen is thus quantum mechanical.  As we
will see, it also depends on {\it a priori} choices of boundary
conditions for the space-time and on a choice of gauge.  Below,
we will concentrate on non-compact, 11 dimensional cosmological
space-times which expand eternally.

\subsec{SUSY and the Holographic Screens}

As explained in the introduction, the connection between local
SUSY and holography comes via the Cartan-Penrose relation between
spinors and null directions. Given a null direction $p^{\mu}$ in
$d$ dimensions, there are always $2^{[d/2] - 1}$ solutions of the
Cartan-Penrose equation

$$p_{\mu} \gamma^{\mu} \psi = 0$$.

Furthermore, the space of all solutions of this equation for
general $p^{\mu}$ can be characterized as the submanifold of all
(projective) spinors satisfying $\bar{\psi} \gamma^{\mu} \psi
\gamma_{\mu}\psi = 0$. The null direction, of $p^{\mu}$ is
determined by this equation, but since the spinor is projective,
not its overall scale.

 A choice of pure spinor completely determines a
null direction and a $(d-2)$ dimensional hyperplane transverse to
it. The orientation of the hyperplane, but not its extent, or
location in space-time, is determined by the nonzero components of
$\bar{\psi} \gamma^{[\mu_1 \ldots \mu_k ]} \psi$ for all $k$. In
light front coordinates where $p^{\mu} \equiv \bar{\psi}
\gamma^{\mu} \psi $ has only a $+$ component, the nonzero
bilinears are $\bar{\psi} \gamma^+ \gamma^{a_1 \ldots a_k} \psi$,
for $0 \leq k \leq d-2$.   These determine the orientation of a
transverse plane.

We interpret this mathematical result physically, by saying that,
given a point in space-time, the choice of a pure spinor at that
point can be thought of as determining the direction and
orientation of the holoscreen on which the information at that
point is encoded.   A hypothetical holoscreen gauge invariance
can thus be thought of as a local transformation that changes a
pure spinor at each point.

In $3,4,6$ and $10$ dimensions, pure spinors can be obtained as
the projective space defined by a linear representation of the
Lorentz group (Dirac, Weyl, Symplectic Majorana, and
Majorana-Weyl).  A general Dirac spinor\foot{I will use the
phrase Dirac spinor to mean a representation of the Dirac
algebra.  In some dimensions, it is appropriate to impose a
reality condition on the spinor and it would generally be
described as a Majorana spinor.  I use the name Dirac in order to
treat all dimensions uniformly.} in dimensions $4,6$ and $10$ can
thus be decomposed into a pair of pure spinors in a canonical
way.  In this decomposition, the two pure Lorentz representations
are related to each other by a discrete element of the Lorentz
group which effects a spatial reflection.  In 3 dimensions, the
minimal Dirac representation is not parity invariant.  If we
double it, in order to represent the parity operation, then all
of these dimensions can be described in a uniform manner.  We
will use the word chiral to refer to the two Lorentz
representations, in any of these dimensions, which are mapped into
each other by parity.

The null direction defined by the vector bilinear of a pure
spinor is always future directed.  However, the same spinor also
satisfies the CP equation for the time reversed null direction.
This is a fundamental ambiguity in the determination of a null
direction satisfying the CP equation, from a given pure spinor. I
will make a convention, choosing the causal direction of the null
vector according to the chirality of the pure spinor
representation.  Then the choice of a Dirac spinor at a point $P$
is equivalent to a choice of a pair of holographic screens, one
in the future, and one in the past of $P$.

 In dimensions other than $3,4,6$ and $10$ there is
no canonical way to associate a general spinor with unique pure
spinors.  We can always find a basis for the spinor
representation consisting entirely of pure spinors. Indeed,
consider two linearly independent null directions, one with
positive and the other with negative time component. The pure
spinor conditions for these two null vectors define two subspaces
of the space of all spinors, with half the dimension of the total
space. Furthermore, since a pure spinor uniquely determines its
null direction (up to a sign), the two subspaces are linearly
independent, and thus form a basis for the entire Dirac spinor
space. However, in general this splitting is not canonical - the
choices of null vectors are arbitrary.  In $3,4,6,$ and $10$
dimensions, there is a canonical way to decompose general spinors
into irreducible representations of the Lorentz group, such that
each spinor in the irrep is pure.

In any spacetime which is a compactification of a ten dimensional
theory, we can decompose a general spinor into pure spinors in a
unique way, so a general spinor at a point in such a space-time
can be thought of as defining the direction and orientation of a
pair of holographic screens on which to project the information
from that point.   As a consequence, a gauge principle that
refers to the ambiguity of choosing a spinor at each point in
spacetime (which is to say, local SUSY) , can be thought of as the
holographic gauge invariance implicit in the work of Bousso.

One of the lacunae in my understanding of this subject is the
lack of an analogous statement about eleven dimensions.  I know
of no way to canonically split a general spinor into pure spinors
in eleven dimensions.  In the discussion below, this will be
dealt with in an unsatisfactory manner.  We will discuss eleven
dimensional Big Bang cosmology, where we will see that the
fundamental variables are past directed pure spinors.  This will
enable us to sidestep the necessity to decompose a general spinor
into pure spinors.

Aficionados of the superembedding formalism
\ref\sorokin{D.P.~Sorokin, {\it Superbranes and Superembeddings},
Phys. Rept. 329, 1, (2002), hep-th/9906142.} will undoubtedly
appreciate the deep connection between what I am trying to do
here and that formalism.  Superembedding is particularly powerful
in eleven dimensions.  I regret that my own understanding of this
formalism is so rudimentary that I cannot exploit its elegance in
the present work. In the super-embedding formalism, the CP
equation often arises as an equation for bosonic spinors which
are world volume SUSY partners of Green-Schwarz variables. From
the point of view of the present paper, one may view it as
specifying the holographic screen for a particular point on the
world volume of the superbrane.

The projective invariance in the definition of pure spinors is a
classical gauge invariance.  It is real or complex, depending on
the nature of the Lorentz irreps in the given space-time
dimension.  When we quantize the pure spinors, we will break most
of this invariance, but there will always be at least a $Z_2$
gauge invariance left over in the quantum theory.   We will see
that this $Z_2$ can be identified with Fermi statistics.  Indeed,
the classical spinors of the CP equation are bosonic.  We will
quantize them as compact bosons - generalized spin operators.
Each such operator will correspond to a new bit of holographic
screen which is added to the Hilbert space describing the
interior of a backward light-cone in a Big Bang space-time, as
one progresses along a timelike trajectory.  Operators
corresponding to independent areas on the screen commute with
each other.  However, the $Z_2$ gauge invariance of the formalism
will enable us to perform a Klein transformation which makes all
the operators into Fermions.  In this new basis for the operator
algebra, the $Z_2$ is just $(-1)^F$.  In some dimensions there
will be a larger group of discrete gauge transformations, which
survives quantization of the CP equation.  I would like to
interpret this as a discrete R symmetry.

\newsec{\bf Holographic cosmology}

In this section I will recall some ideas about holographic
cosmology that were presented in \ref\tbmill\
\ref\bfmcosmo{T.~Banks, W.~Fischler, {\it M-theory Observables
for Cosmological Space-Times}, hep-th/0102077.} , and extend them
by choosing the fundamental variables of the theory to be
quantized versions of the pure spinors of Cartan and Penrose. The
formulation of quantum cosmology that I will present is
Hamiltonian, and therefore necessarily gauge fixed. Thus, much of
the discussion of local SUSY and holographic gauge invariance
will not be immediately apparent.   The reason for this is that
the fundamental clue about how to quantize the system is the
BHFSB entropy bound.   This refers to physical quantum states of
the system, and so the formalism I present will be in a fixed
gauge.  Nonetheless, we will see evidence for the physical
components of the gravitino field in the formalism.

\subsec{Prolegomenon to a Holographic Theory of Space Time}

The treatment of cosmology in string theory has, for the most
part, been an exercise in effective field theory. Many
cosmological solutions of the equations of low energy
perturbative string theory can be found, but like most time
dependent solutions of Einstein's equations, they contain
singularities, in this case Big Bang or Big Crunch singularities.
This indicates the necessity for a more profound approach to the
problem. Fischler and Susskind provided a fundamental new insight
into cosmology, and Big Bang singularities, by trying to impose
the holographic principle in a cosmological
context\ref\lenwilly{W.~Fischler, L.~Susskind, {\it Holography
and Cosmology},hep-th/9806039. }. This work was followed by
Bousso's construction of a completely covariant holographic
entropy bound\raph\ .

I believe that the work of F(ischler) S(usskind) and B(ousso)
provided us with the foundations of a quantum theory of cosmology.
There are three important principles that are implicit in the work
of FSB:

\noindent 1. The holographic principle is consistent with the idea
of a {\it particle horizon}, a notion which we generally derive
from local field theory.  More generally, it is consistent with
the idea that a {\it causal diamond} in space-time contains an
operator algebra that describes all measurements, which can be
performed within this diamond.  In many cases, the holographic
principle implies that the dimension of this operator algebra is
finite. In this case there is a unique Hilbert space
representation of the algebra.

\noindent 2. In particular, the interiors of backward light cones
in a Big Bang spacetime must have finite operator algebras.
Furthermore, the FSB entropy bound implies that the dimension
decreases (apparently to zero) as we go back to the Big Bang
singularity. These results lead one to conjecture that instead,
the (reverse) evolution stops when the Hilbert space has some
minimal dimension.  They also lead one to some guesses about the
fundamental formulation of quantum cosmology, which I will sketch
below.

\noindent 3.  Within the semiclassical approximation the
holographic principle is compatible with
F(riedman)-R(obertson)-W(alker) cosmology at early times if and
only if the stress tensor satisfies the equation of state
$p=\rho$, with the entropy density related to the energy density
by $\sigma \propto \rho^{1\over 2}$ \foot{The latter condition
means that the homogeneous modes of minimally coupled scalars
{\it do not} satisfies the requirements of holography.}. This is
a peculiar new form of matter.  Fischler and I
\ref\holocosmo{T.~Banks, W.~Fischler, {\it An Holographic
Cosmology}, hep-th/0111142.} gave a heuristic picture of such a
system as a \lq\lq dense fluid of black holes \rq\rq , but a more
precise quantum description still eludes us.

I would like to sketch the outlines of a quantum cosmology based
on these principles.  This sketch is an update of ideas presented
in \tbmill\ref\bfmcosmo{T.~Banks, W.~Fischler, {\it M-theory
Observables for Cosmological Space Times}, hep-th/0102077.} . It
is still less than a full dynamical quantum theory of space-time.
In presenting it, I have used the following strategy. I utilize
space-time concepts to motivate quantum mechanical constructions.
Eventually, one would like to turn everything around, and present
a set of purely quantum axioms from which we derive a classical
space-time.  The reader should keep in mind the dual purpose of
this discussion and, as it were, try to read every argument both
backwards and forwards .

Consider a time-like trajectory (perhaps a geodesic) in a Big Bang
space-time, and a sequence of backward light cones whose tips end
on this trajectory.  The FSB bound implies that the Hilbert space
describing all measurements in the interior of each of these light
cones is finite dimensional. Let us define the entropy to be the
logarithm of this dimension (it is the entropy of the maximally
uncertain density matrix on this Hilbert space). Let us for the
moment restrict attention to a period in which the universe is
expanding. Then the entropy decreases as we follow the trajectory
back to the Big Bang.

The concept of particle horizon means that each of these Hilbert
spaces should have a self contained description of all of the
physics that goes on inside it.  That is, there should be a
sequence of unitary transformations describing time evolution
inside each backward light cone, without reference to any of the
larger light cones.  On the other hand, the dynamics in a large
light cone should be restricted by consistency with earlier light
cones in the sequence.  The reason that I insist on a sequence of
unitaries, rather than a continuous one parameter family is that
the system in any one light cone is finite dimensional.  A finite
system can have a continuous time evolution if it is in contact
with an external classical measuring apparatus, but, because of
the time-energy uncertainty relations (whose precise form depends
on the spectrum of the finite system) it does not make sense to
talk about infinitely precise time resolution as a measurement
performed by the system on itself. A more fundamental reason for
discreteness will be discussed below.

This suggests, that as time goes on and the particle horizon
expands, more and more precise time resolution becomes available.
Thus, the time intervals between unitary transformations in the
sequence should not be thought of as defining the Planck time.
Instead, I insist that they define time slices in which the FSB
area increases by some minimal amount (to be quantified below).
Call the sequence of Hilbert spaces ${\cal H}_n$. $\hn$ has
dimension $K^n$, and we think of it as defining the Hilbert space
inside a backward light cone whose holographic screen has FSB area
$4 n {\rm ln} K$ in Planck units. In each $\hn$ there is a
sequence of unitary transformations $U_n (k)$ for $n \geq k \geq
0$. One further assumes that $\hn = H \otimes {\cal H}_{n-1}$,
where $H$ is a $K$ dimensional space.  The maps $U_n (k)$ are
required to factorize in a manner compatible with this
concatenated tensor factorization of the Hilbert space.  For
example, for every $n$ and $k$, $U_n (k)$ for $k < n$ is a tensor
product of $U_{n-1} (k)$ and a $K$ dimensional unitary
transformation on $H$.  Below, we will choose the number $K$ in a
natural way that depends on the dimension of space-time.

This definition gives us some idea of how much time is represented
by each unitary evolution in the sequence.  An area $4n\ {\rm ln}\
K$ in $d$ space-time dimensions, allows the creation of black
holes of energy of order $(4n\ {\rm ln}\ K)^{(d-3)\over (d-2)} $.
The inverse of this energy is the maximum time resolution that
such a system can have.   On the other hand, if we make some other
assumption about the state of the system, we may have less time
resolution than this.  Thus, we can begin to see a correlation
between the space-time geometry and the matter content of the
system.

We also see the fundamental reason for discreteness in these
equations.  The FSB areas of backward light cones in a Big Bang
space-time are quantized because they refer to the logarithms of
Hilbert space dimensions.

 So far of course we have defined much less than a full
space-time. To go on, we need to consider neighboring time-like
trajectories, and we must introduce the dimension of space-time.
To do this, introduce a $d$ dimensional cubic lattice, and assign
Hilbert spaces and unitary operators to each vertex of the
lattice.

There are several disturbing things about this (as far as I can
see) unavoidable introduction of dimensions.  The is that we
believe that we can define cosmologies in string theory, that
interpolate between spaces of different dimension. For example,
the Kasner cosmologies studied in \ref\bmfbm{ T.~Banks,
W.~Fischler, L.~Motl, {\it Duality vs. Singularities}, JHEP 9901
(1999) 019, hep-th/9811194; T.~Banks, L.~Motl, {\it On the
Hyperbolic Structure of Moduli Space With Sixteen SUSYs}, JHEP
9905:015,1999, JHEP 9905:015,1999.} can interpolate between
heterotic strings on tori and 11D SUGRA on K3 manifolds. It is
not clear to me that this is a difficulty.  We are not describing
local field theories here, and our description might be valid in
all regions of moduli space, even though defined with respect to
one.  What is certain, is that all of these dualities involve the
nontrivial topology of the compactification manifold. We can for
the moment restrict our attention to describing the noncompact
part of space, with the compact parts described by the structure
of the spectrum of states in Hilbert space.  However, there is
obviously much to be understood about this question. In the
present paper I will restrict attention to non-compact $11$
dimensional cosmologies.

The second disturbing aspect of our construction will be an
asymmetry between space and time. It is intrinsic to our
formulation of the problem in terms of time evolution in Hilbert
space (rather than some sort of path integral formalism). We have
chosen a rather particular gauge, in which every point on a time
slice has a backward light-cone with equal FSB area.  One could
make different choices, but none would be gauge independent.  This
is the famous problem of time in Quantum General Relativity. No
physical Hamiltonian of a general covariant theory can be gauge
independent, since the choice of time evolution is a choice of
gauge.  Only in space-times with a fixed classical asymptotic
boundary can we imagine a gauge independent choice of Hamiltonian,
because we insist that the group of gauge transformations consists
only of those which fall off at infinity . We will introduce the
asymmetry between space and time into our notation by labelling
points in the lattice by a $d$ vector of integers $(t, {\bf x})$.

Now we have to address the question of how the Hilbert spaces and
time evolution operators corresponding to different points on the
lattice, are related to each other.  It is here that the formalism
parts company with a lattice field theory like system, where each
point should have independent degrees of freedom.  In fact, since
we are associating the observables with experiments done in the
backward light cone of the point, there should be a large degree
of overlap between nearest neighbors. Indeed, we defined the
smallest time difference by insisting that the Hilbert space at
time $n$ have only $K$ times as many states as that at time
$n-1$.  If $K$ were $2$, this would be the minimum increase
compatible with the notion that the new particle horizon has some
independent degrees of freedom in it that were not measurable in
the old one. Similarly we will require a maximal overlap for
nearest neighbor points on the lattice.  That is, the Hilbert
spaces $\hn (x) $ and $\hn (x + \mu)$ should each factorize as
\eqn\factor{\hn (x) = H(x) \otimes O(x,x+\mu )\ \ \hn (x+\mu ) =
H(x+\mu ) \otimes O(x, x+\mu ), }  where for each $y$, $H(y)$ is a
$K$ dimensional space.  The relation between nearest neighbor
Hilbert spaces will be made more explicit below, once we
introduce a set of generators for the operator algebras in each
space.

We will choose $K$ in a manner motivated by our remarks about the
connection between supersymmetry and holography.  Let $S_{\alpha}$
transform in the Dirac spinor representation of the Lorentz group
$Spin(1,d-1)$.  The details of the construction will depend
somewhat on the properties of spinors in various dimensions, so I
will restrict attention to d=11.   We will insist that the spinor
be pure, that is , that $\bar{S} \gamma^{\mu} S \gamma_{\mu} S =
0$. Such spinors have $16$ independent real components.   In the
quantum theory, they will be quantum operators, $S_a$, $a=1\ldots
16$ .  We also restrict attention to past directed pure spinors -
the associated null vector is past directed.  In choosing to
describe the pure spinor in terms of only sixteen variables, we
have chosen a gauge for local Lorentz gauge symmetry.  In
principle, one could keep $32$ components and a local symmetry
which allowed us to reduce to $16$.  However, the Lorentz
connection would have to be a constrained variable, in order not
to introduce new degrees of freedom into the system.  We are
aiming toward a completely gauge fixed Hamiltonian description of
our cosmology.  In such a description a constrained variable has
been solved for in terms of the physical dynamical variables in
that gauge.  Below, we will introduce a mapping $\Psi$ between the
operator algebras in Hilbert spaces at different points on the
lattice.  In particular, that mapping will relate the spinor
basis at one point to that at another. $\Psi$ implicitly contains
the gauge fixed Lorentz connection.

We have seen that, classically, a pure spinor determines a past
directed null direction.  We think of the physical interpretation
of this null direction in terms of two holographic screens for an
observer traveling along the timelike trajectory between $(t,{\bf
x})$ and $(t+1, {\bf x})$ .   The physics inside the backward
light cone of the observer at these two points, can be projected
onto a pair of holographic screens, both in the past of the tips
of the light cones. In a bulk geometrical picture, the information
that is not contained in the smaller screen can be communicated
to the observer at some point, P,  on his trajectory between the
tips of the two light cones. The new pure spinor that we add to
the system may be thought of as the instruction for building the
new piece of the holographic screen, on which the information at
P is to be projected.  Of course, this classical language can
have only a poetic meaning at the time scales on which we are
making our construction.

It is important to note that the paragraph above contains the
answer to the question of where and how large the holographic
screen is.  If we assume that the quantum formalism will, in the
limit of large Hilbert spaces, indeed determine a classical
geometry consistent with the words we have been using, then the
bit of holographic screen that is added by the operator $S_a
(t,{\bf x})$ is located on the FSB surface of the backward
light-cone from $(t+1, {\bf x})$ , and has area $4\ {\rm ln}\
256$ in Planck units.  The null vector which would specify
precisely where on that screen this particular variable is, is
the bilinear current constructed from this pure spinor.  It is a
quantum operator, and so only describes probability amplitudes
for the bit of screen to be at specific points on the FSB surface.
The FSB surface itself is constructed out of all the spinors in
the Hilbert space ${\cal H}_{n+1}$, so its quantum fluctuations
are small in the limit of large area.

As anticipated above, we will build the Hilbert space ${\cal H}(t
+ 1, {\bf x}) $ by adding operators $\hat{S}_a (t+1, {\bf x})$ to
the Hilbert space ${\cal H}(t, {\bf x})$. These will commute with
all of the operators in the latter space.   The defining relation
for a pure eleven dimensional spinor is invariant under real
projective transformations of the spinor. We will break this
invariance in the quantum theory.  We will make the optimistic
assumption that we do not need any other variables to describe
quantum gravity.  Apart from simplicity, this assumption is
motivated by the fact that (when singularities are properly
treated), all degrees of freedom in the standard model can be
viewed as parts of supergravity.  This is one of the lessons of
String Duality.

To quantize the CP spinors, we postulate that \eqn\acr{[\hat{S}_a
, \hat{S}_b ]_+ = 2\delta_{ab}.}  Up to normalization, this is the
unique ansatz that gives a finite dimensional Hilbert space, and
is invariant under the $SO(9)$ group of rotations that leave the
null vector invariant.  These postulates break the projective
invariance except for a factor of $(- 1)$.  We will treat the
latter factor as a $Z_2$ gauge transformation, which will
eventually be seen as Fermi statistics.  The fact that the
classical projective gauge symmetry of the CP equation is broken
down to $Z_2$ has to do with the fact that our spinor carries
information about the conformal factor of the spacetime geometry,
as well as its causal structure. Indeed, the commutation relations
determine the dimension of the new Hilbert space, and thus the
area of the new holographic screen.  The logarithm of the
dimension of the new Hilbert space increases by $8\ {\rm ln}\ 2$,
which corresponds to an increase in area (Planck units) of $32\
{\rm ln}\ 2$.

We can now turn the $\hat{S}_a$ into Fermions, by defining $S_a =
(-1)^F \hat{S}_a$, where, $(-1)^F$ is the product of all of the
previous $S_a$ operators (note that the number of these operators
is always even) . In other words, we start with the irreducible
representation of the Clifford algebra. This defines the smallest
possible Hilbert space at the moment of the Big Bang. Then we
build successive Hilbert spaces along a given timelike
trajectory, by tensoring in one more commuting copy of the
minimal Clifford representation. We then do a Klein
transformation to present the full algebra as a larger Clifford
algebra.  The Klein transformation is a $Z_2$ gauge
transformation, which is the quantum remnant of the projective
invariance of the Cartan-Penrose equation.   It is Fermi
statistics of the Klein transformed operators.  Note that all
operators transforming in integer spin representations of the
Lorentz group, will be even functions of the $S_a$, so the
connection between spin and statistics is built into the
formalism.

To recapitulate, the quantum description of the causal pasts of a
sequence of points along a given timelike trajectory in a Big
Bang cosmology is described by a sequence of Hilbert spaces
${\cal H}_n$.   The operator algebra of the $k$th Hilbert space
is the Clifford algebra generated by operators $S_a (n)$ with $1
\leq n \leq k$:

 \eqn\onehor{[S_a (n), S_b (m)]_+ = \delta_{ab}\delta_{mn}}
The operator $S_a (n)$ in each Hilbert space may be identified
with the operator with the same labels in any other Hilbert space.
We will see later that this identification may be viewed as a
gauge choice for the discrete analog of local SUSY.

Dynamics is defined by a sequence of unitary transformations, $\{
U(n) \}$, in each Hilbert space, satisfying a simple compatibility
condition, which will be discussed below.  In principle, we could
introduce a continuous unitary groupoid $U(t,t_0 )$ such that the
unitary transformations in the sequence could be viewed as the
values of $U(t_n, 0 )$ at a sequence of times.  In this way the
formalism becomes that of ordinary quantum mechanics, with a time
dependent Hamiltonian,  but changes in the groupoid, which do not
change the values at the special times $t_n$, should be viewed as
physically equivalent.

The sequence in ${\cal H}_k$ has $k$ steps, and should be thought
of as the evolution operators over the time steps determined by
the sequence of points on our timelike trajectory.  The
fundamental consistency condition is that the operator $U_k (n)$
in ${\cal H}_k $with $n < k$ should be a tensor product of the
($k$th copy of) $U_n (n)$ with an operator that depends only on
the $\hat{S}_a (m)$ with $ m > n$.   Thus \eqn\consist{ U_k (n) =
U_n (n) V_k (n),} where $V_k (n)$ is a function only of $S_a (m)$
with $m >n$.  We will impose the $Z_2$ gauge invariance on all of
these unitaries, so that they are even functions of the
fundamental variables, and we can ignore the distinction between
the hatted and bare headed variables.  Note also, that in writing
the last equation we have used the same notation for the operator
$U_n (n)$ and the copy of this operator in every ${\cal H}_k $
with $k > n$.

These rules define a quantum system, which is compatible with the
notion of particle horizon in a Big Bang cosmology.  The Hilbert
space ${\cal H}_k$ describes all measurements that can be done
inside the particle horizon at time $t_k$, in a manner compatible
with the fact that measurements inside earlier particle horizons
commute with measurements that can only be made at later times.
Each particle horizon has its own time evolution operator, but
the evolution operators at early times, agree with those in
previous particle horizons, in their action on those variables
that are shared between the two systems\foot{From here on I will
stop insisting that the shared operators are really copies of the
operators at earlier times. The reader will have to supply this
pedantry by himself.  Its importance will be apparent when we
discuss copies of operators associated with other timelike
trajectories.}

The system is also compatible with the holographic principle in
that we will identify the dimension of the Hilbert space with the
area of the FSB surface on the past light cone\foot{This
identification is only appropriate for the past light cones on
trajectories in an eternally expanding universe. In contracting
universes, the FSB area can sometimes decrease as one goes into
the future. It must then be interpreted as the entropy of a
density matrix more pure than the uniform probability density. The
interpretation of this is that the assumed spacetime geometry and
matter content is a very special class of states of the system.
More generic initial conditions at times before the FSB area
began to decrease would have led to a different spacetime in
these regions.  Of course, the latter statement could also be
made about, {\it e.g.}, the future evolution of an expanding
matter dominated FRW universe. However, in this case we can still
imagine exciting a more general configuration in the future by
creating lots of black holes. In contracting regions, certain
possible excitations of the system at early times are ruled out
by the assumption that the geometry behaves in a particular
classical manner.  In this connection note that if we examine
contracting FRW universes with matter with equation of state
$p=\rho$, corresponding to a maximal entropy black hole fluid,
then the FSB area of backward light cones always increases.  It
is only the assumption that low entropy systems with soft
equations of state persist into contracting regions that leads to
the phenomenon of decreasing area. }. This statement does not have
much content until we enrich our system and show that it does
have a spacetime interpretation.

Indeed, the conditions we have stated so far are very easy to
satisfy, and most solutions do not resemble spacetime in any
obvious way.   What is missing is the notion that the new degrees
of freedom that come into a particle horizon ``come from other
points in space".   To implement this, we return to our hypercubic
eleven dimensional lattice with points labeled $(t, {\bf x})$ and
Hilbert spaces ${\cal H} (t, {\bf x})$.  The sequence of Hilbert
spaces at fixed ${\bf x}$ has all the properties we have
described above.

To understand the geometric interpretation of , {\it e.g.}, the
Hilbert space ${\cal H} (t, {\bf x + e_1}$, where ${\bf e_1}$ is
some unit lattice vector, introduce a time slicing of our Big
Bang spacetime by the rule that the past light cone of every
point on a time slice has equal FSB area (for FRW these are just
slices of cosmic time, but a single unit in $t$ does not
correspond to a fixed unit of cosmic time, but rather a fixed
unit of FSB area) .  Now, starting at a point labeled ${\bf x}$
on a fixed time slice, choose a spacelike direction on the slice
and move along it to a new point, labeled ${\bf x + e_1}$.  The
intersection of the past light cones of these two points has an
almost everywhere null boundary but is not a full light cone.
Choose the point (for very small distance between the two points
on the time slice, it will be unique) inside the intersection
whose past light cone has the largest FSB area, and call this the
FSB area of the intersection . The point ${\bf x} + {\bf e_1}$ is
chosen such that the FSB area of the intersection is smaller than
the FSB areas of the causal pasts of ${\bf x}$ and ${\bf x} +
{\bf e_1}$, by precisely the fundamental unit. Now proceed to do
the same in the negative $e_1$ direction and in $9$ other locally
independent directions. Then repeat the same procedure for each
of these new points and so on {\it ad infinitum} (for this paper,
we restrict attention to spatial topologies, which are trivial and
extend to infinity in all dimensions).   Repeat the same for each
time slice.  This picture motivates our lattice of Hilbert spaces.

The crucial step now is to introduce maps between a tensor factor
of the operator algebra (equivalently, the Hilbert space, since
everything is finite dimensional) in ${\cal H} (t, {\bf x}) $ and
that in ${\cal H} (t, {\bf x + e_i})$ for every (positive and
negative) direction.  The common factor Hilbert space has
dimension smaller by a factor of ${1\over 256}$.  Equivalently we
can think of this as a relation, which defines a copy of the
generators of the algebra at ${\bf x}$ in the Hilbert space at
${\bf x + e_i}$.

\eqn\connection{S_a (t_i; t , {\bf x}; {\bf x + e_1}) = \Psi_{ab}
(t_i , t_j; {\bf x, x+e_1}) S_b (t_j , {bf x + e_1})}

The labels $0\leq t_i \leq t ,0\leq t_j \leq t$ on the operators,
remind us that the Hilbert space ${\cal H} (t, {\bf x})$ contains
operators that have been copied from the Hilbert spaces at all
previous times.  The map $\Psi$ is part of the definition of the
dynamics of this quantum spacetime.   It is subject to a large
number of constraints.  Viewed as a matrix on the $256 t$
dimensional space of $S$ components, it should have rank $256
(t-1)$.  Precisely $256$ generators of the algebra at ${\bf x}$
should have vanishing representative in the nearest neighbor
Hilbert space. Furthermore, the different $\Psi$ maps at different
points of the spacetime lattice must all be compatible with each
other.

A much stronger set of constraints comes from requiring that the
unitary transformations $U(t_k, 0)$ in each Hilbert space be
compatible with each other after application of the map $\Psi$.
This is a system of mutual compatibility constraints between the
$\Psi$ maps and the unitary transformations. Indeed, one is
tempted to conjecture that any lattice of Hilbert spaces, $\Psi$
maps and unitary transformations satisfying all of these axioms
should be viewed as a consistent quantum mechanical description
of a Big Bang cosmology.  I am not prepared to make such a bold
conjecture at this time.  Many examples of solutions to these
constraints will have to be discovered and worked out before we
can hope to understand this formalism, and whether it needs to be
supplemented with additional axioms.

The bilateral relations between nearest neighbor Hilbert spaces
on the lattice, enable us construct copies of subalgebras of the
operators in any Hilbert space, inside the operator algebra of
any other. For a pair of points on the lattice, this
correspondence will be path dependent.   Thus, in some sense, the
fundamental dynamical variables in the theory are the path
dependent objects

\eqn\gravitino{S^{\Gamma}_a (t, {\bf x} ; t^{\prime}, {\bf
x^{\prime}}).}

In words, this is the copy of $S_a (t, {\bf x})$, in ${\cal H}
(t^{\prime} , {\bf x^{\prime}} ) $ obtained by concatenating the
$\Psi$ maps along the path $\Gamma$ between the two points. The
$\Psi$ map gives us a special case of these variables for the
minimal path between nearest neighbor points. Thus, $S_a (t, {\bf
x} ; t, {\bf x + e_1}) $ can be thought of as a discrete analog
of the gravitino field $\psi_{a \mu} de_1^{\mu }$ integrated
along the link between two nearest neighbor lattice points.

We now see that the simple mapping between the operator algebras
at different times, at the same spatial point, can be viewed as a
gauge choice for the time component of the gravitino field. We do
not yet have any evidence that this formalism reduces to some
kind of classical field theory in limiting situations, but it
seems likely that if it does, that field theory will be locally
SUSic.

\subsec{Discussion}

The system that we have been discussing bears some resemblance to
a lattice quantum field theory.  This is both misleading, and
suggestive.  It is misleading because the fields at different
points of the lattice at the same time do not (anti)-commute with
each other. Their commutation relations are complicated and
depend on the choice of $\Psi$ mappings. This choice is part of
the specification of the dynamics of the system. Note further
that if the true connection between geometry and quantum
mechanics is to be extracted from the entropy/area relation, the
space-time geometry will not be that of the lattice.  The lattice
does serve to specify the topology of the spacetime.

The relation to lattice field theory is however suggestive of the
possibility that in some dynamical circumstances, suitable
subsets of the variables of this system might behave like quantum
fields.

\newsec{\bf What is To Be Done?}

The answer to this question is of course: ``Almost everything".
More specifically, the most urgent problem is to find one example
of a solution to the constraints postulated above, and show that
in the large time limit it has an approximate description in
terms of quantum fields in curved spacetime.   The obvious case
to start with is that of a homogeneous isotropic universe.  That
is, every sequence of Hilbert spaces ${\cal H} ({\bf x}, t)$ has
the same set of unitary maps $U(s,s-1 )$ for $s \leq t$.  There is
essentially a single $\Psi$ mapping, which must be consistent with
the unitary dynamics.  This problem is still complicated enough
that no solutions have been found as yet.

With W. Fischler, I have conjectured a possible solution,
corresponding to a homogeneous spatially flat universe with
equation of state $p=\rho$.   Write $U(s,s-1) = e^{H(s)}$ and
expand $H(s)$ in powers of the Fermion operators.  In particular,
there will be a quadratic term

\eqn\hquad{H_2 (s) = \sum_{p,r < s} S_a (p) h(s|p,r) S_a (r)}

Now, for each $s$, let $h(s | p, r)$ be a random antisymmetric
matrix, chosen from the Gaussian ensemble.  It is well
known\ref\kwd{F.J.~Dyson, J. Math. Phys. 3, 140 (1962), 3, 1191,
(1962), 3, 1199, (1962) ; V.~Kaplunovsky, M.~Weinstein, {\it
Space-Time: Arena or Illusion?}, Phys. Rev. D31, 1879, (1985),
use this connection in a manner similar to what is presented
here.} that large Gaussian random antisymmetric matrices have a
spectral density that behaves linearly in a larger and larger
region around zero eigenvalues.  Thus, for large $s$, the
spectrum of the time dependent Hamiltonian $H_2 (s)$has a
universal behavior that looks like that of a system of free
massless Fermions in $1+1$ dimensions.  Furthermore, with the
exception of a single marginally relevant four Fermion operator
(the analog of the Bardeen-Cooper-Schrieffer operator), this low
energy spectrum will not be disturbed by higher order polynomials
in Fermions. The universal behavior of the spectral density is
shared by a large class of random Hamiltonians for the Fermion
system.

Thus, for large $s$, although we are dealing with a problem with
time dependent Hamiltonian, we approach a system whose spectral
density becomes time independent and satisfies the energy/entropy
relation $\sigma \sim \sqrt{\rho}$ of a $p=\rho$ fluid.  Note
that, in the hypothetical translation of this physics into a
spacetime picture, the energy of this system at time $t$ would be
interpreted as the energy density at the tip of the backward light
cone, $(t, {\bf x})$.

On the other hand, because for each $s$ we make an independent
choice of random Hamiltonian, there is no sense in which the
quantum state of the system settles down to the ground state of
any given Hamiltonian, even after time averaging.   All of the
degrees of freedom of the system remain permanently excited. The
density matrix of the system is completely random, maximizing the
entropy, but certain energetic properties become smooth and
universal for large $s$.  To prove that this system satisfies our
axioms, one would have to exhibit a $\Psi$ mapping compatible
with this prescription for time evolution. This has not yet been
done.

Our discussion has been restricted to Big Bang cosmological
spacetimes.  I have emphasized elsewhere\ref\hetero{T.~Banks,
{\it A Critique of Pure String Theory}, Talk given at Strings
2002, Cambridge, UK, July 2002; {\it A Critique of Pure String
Theory or Heterodox Opinions of Diverse Dimensions}, manuscript
in preparation.} that one should expect the fundamental
formulation of quantum gravity to depend on the asymptotic
geometry of spacetime.  Gravity is not a local theory, once one
goes beyond the realm of classical geometry (where the degree of
non-locality can be controlled by the choice of initial
conditions of the classical solution), and its fundamental
formulation has every right to depend on the boundary conditions
.  Nonetheless, the considerations of the present paper suggest
the possibility of a more local, but perhaps gauge dependent
formulation, as has been advocated by Susskind. Again, the idea
is that the choice of holographic screen is a gauge artifact.
In asymptotically flat or AdS spacetimes, it may be convenient
and elegant to place the screen at infinity, but there may also
be other gauges where the same information is mapped onto a
collection of local screens.  Our formulation of quantum
cosmology has fixed a particular gauge defined by equal area time
slices.

In asymptotically flat spacetime, the causal diamond formed by
the intersection of the causal past of a point with the causal
future of point in that causal past, has finite FSB area. Thus,
one can imagine assigning finite dimensional operator algebras to
causal diamonds and trying to imitate the formalism of this paper.
Any given finite dimensional algebra would be embedded in a
sequence of algebras corresponding to larger and larger causal
diamonds.  The limiting Hilbert space would be infinite and the
limiting time evolution operator in this space would approach the
scattering matrix.   Again, the formalism would be constrained by
the requirement of consistency with many partially overlapping
sequences corresponding to nested causal diamonds centered around
different points in space.

In asymptotically flat spacetime we expect to have an exact
rotational symmetry.     Thus, it makes sense to choose
holographic screens which are spherically symmetric. One would
want to represent the pure spinor operators for a finite causal
diamond as something like elements of a spinor bundle over a
fuzzy sphere in order to have a formalism which preserves
rotational invariance at every step.   Furthermore, the explicit
breaking of TCP invariance which was evident in our treatment of
Big Bang cosmologies, should be abandoned.  Each causal diamond
of FSB area $K^N$ should have a sequence of unitary operators
$U(t_k , - t_k)$ for $ 1 \leq t_k \leq N$, which commute with an
anti-unitary TCP operator.  In the limit as $N\rightarrow\infty$,
$U(t_N , - t_N)$ would become the S-matrix.  An important aspect
of this limit is that the finite $N$ fuzzy sphere should become a
conformal sphere as $N\rightarrow\infty$, in order to obtain a
Lorentz invariant S-matrix (the Lorentz group is realized as the
conformal group of null-infinity).

In asymptotically AdS spacetime, things are more complicated. The
causal past of a point includes all of AdS space prior to some
spacelike slice.  Thus if we try to construct the causal diamond
corresponding to a pair of timelike separated points, it becomes
infinite when the timelike separation is of order the AdS radius,
and the backward and forward light-cones of the two points
intersect the boundary of AdS space before intersecting each
other. This suggests that there should be a sequence of finite
dimensional operator algebras which cuts off at some finite
dimension of order $e^{R_{AdS}^{(d-2)}}$.   Since we already have
a ``complete" formulation of the quantum theory of AdS spacetimes,
it would seem to be a good strategy to search for such a sequence
of nested operator algebras within the Hilbert space of conformal
field theory. This would be a new approach to the puzzle of how
local data is encoded in the CFT.

\listrefs

\bye